\documentclass[10pt,letterpaper,twocolumn]{article} 

\usepackage{ol2}
\usepackage[draft]{hyperref}
\usepackage{amsmath}

\begin{document}

\twocolumn[ 

\title{Active stabilization of a Michelson interferometer at an arbitrary phase with sub-nm resolution}

\vspace{-10pt}


\author{Davide Grassani$^1$, Matteo Galli$^1$, Daniele Bajoni$^{2,*}$ }

\address{
$^1$Dipartimento di Fisica, Universit\`{a} degli Studi di Pavia, via Bassi 6, Pavia, Italy
\\
$^2$Dipartimento di Ingegneria Industriale e dell'Informazione, Universit\`{a} degli Studi di Pavia, via Ferrata 1, Pavia, Italy
\\
$^*$Corresponding author: daniele.bajoni@unipv.it
}
\vspace{-10pt}
\begin{abstract}

We report on the active stabilization of a Michelson interferometer at an arbitrary phase angle with a precision better than one degree at $\lambda = 632.8$ nm, which corresponds to an optical path difference between the two arms of less than 1 nm. The stabilization method is ditherless and the error signal is computed from the spatial shift of the interference pattern of a reference laser, measured in real-time with a CCD array detector. We discuss the usefulness of this method for nanopositioning, optical interferometry and quantum optical experiments.

\end{abstract}

\ocis{120.3180, 120.5050, 260.3160}

 ] 
\vspace{-30pt}

The active stabilization of an optical interferometer is of fundamental importance in a number of optical experiments including holographic recording \cite{Frejlich,Barbosa,Freschi}, phase shifting interferometry \cite{Gnauck} and deep UV lithography \cite{Lith}.
In most of these applications, the differential drift in the two arms of the interferometer caused by environmental noise must be controlled to a very high degree of accuracy, requiring the real-time compensation of the optical path difference between the two interferometer's arms with sub-nanometer precision. Moreover, the ability to adjust and keep the phase difference to an arbitrary angle is also a crucial requirement in a number of applications such as, for instance, high precision metrology \cite{metrology1,metrology2} or quantum optics experiments \cite{Smithey}. 

\begin{figure}[h!]
\centerline{\includegraphics[width=0.9\columnwidth]{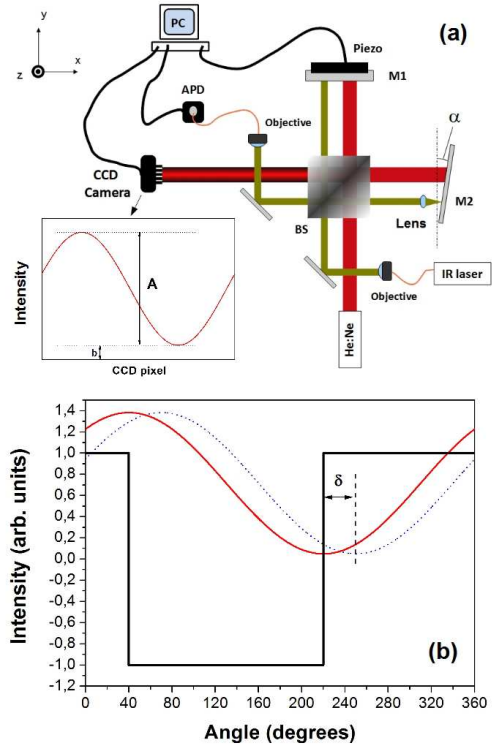}}
\caption{(a) Interferometer set-up. (b) Example of an actual (blue dotted) output waveform on the CCD camera, compared to the desired (red) waveform with the relative step function. $\delta$ is the error with respect to the target angle.}
\vspace{-15pt}
\label{Fig1}
\end{figure} 

A commonly used stabilization method is based on an error signal generated from the output of the interferometer and fed back to a phase compensating mechanism on one of its two arms, typically consisting of a piezoelectric actuator moving one of the mirrors.
One possible solution for phase control is to consider the error signal as the difference between the two outputs of a Mach-Zehnder \cite{Jackson} or a Michelson interferometer \cite{Gray}.  This two outputs are out of phase by $180^\circ$ so the error signal is proportional to the cosine of the interferometer's phase and it is used to lock the device at an integer multiple of $90^\circ$. The periodicity of the error signal can however lead to fringe skipping for perturbations greater than $90^\circ$, and some phases cannot be stabilized due to the periodicity of the error amplitude.

A common method to achieve continuous phase locking is to employ a small modulation of the movable interferometer's arm, known as dither \cite{Freschi2}. Dithers allow to estimate the first derivative of the sinusoidal interferometer's response, thus yielding stabilization at an arbitrary phase, including extremants. The drawback of this method is the modulation itself, which can be considered as an additional source of noise. There is also a trade-off between the dither intensity and the efficiency in the detection of the positioning error, so that in practice it is very difficult to obtain phase stabilization to better than several degrees. Moreover, this method cannot be employed when the desired acquisition rate is faster than the modulation frequency and, for large enough dithers, the hysteresis of the piezoelectric compensator and the reduction of fringe visibility must be taken into account.
Other approaches to generate an error signal that is linear with the phase difference between the two interfering beams involve the use of dispersive elements inside the arms \cite{Krishnamachari} or polarization optics \cite{Jotzu} to retrieve the absolute phase: in both cases the additional optics impose limits on the physical properties of the controlled beam (i.e. its wavelength or polarization state), and both methods are strongly reliant on a precise initial calibration of the system.

In this paper we present a simple method to stabilize an optical interferometer with a positioning accuracy better than one nanometer, without the need of any signal modulation, calibration techniques or additional optics inside the interferometer (with the exception, in some cases, of a single lens, as discussed below). In addition, the presented method allows one to lock the phase of the interferometer at an arbitrary angle with a standard deviation limited by the positioning precision of the piezo controller (0.8 nm in our case). The working principle is illustrated in fig. \ref{Fig1}. The set-up is a simple Michelson interferometer (a Mach Zehnder would work as well), where an intensity- and phase-stabilized laser source of wavelength $\lambda_P$ nm is used as a reference beam (large red continuous line in the figure). The first mirror M1 of the interferometer is mounted on a piezoelectric actuator moving in the $y$ direction (controlling the optical path difference between the two arms), while the second mirror M2 is fixed and slightly tilted by a small angle $\alpha$ around the $z$-axis. This results in the formation of vertical spatial fringes at the exit port of the interferometer, which are detected using a CCD camera.

By restricting the CCD sensitive region to one fringe only in the beam, the signal measured by the sensor will be (inset of Fig. \ref{Fig1} (a)):
\begin{equation}
\label{signal}
f\left(\vartheta,\varphi\right)= b+A\sin\left(\vartheta - \varphi\right)
\end{equation}
where the independent variable $\vartheta$ is discretized by the horizontal dimensions of the CCD pixels (7 $\mu$m in our case), and the factor $A$, which is proportional to the laser intensity, is discretized by the CCD analog to digital resolution. The flat background signal $b$ accounts for the imperfect fringe contrast as well as for the CCD dark current. While $\varphi$ is the interferometer's phase, which is directly linked to the optical path difference $\Delta L$ as $\varphi=720^\circ*\frac{\Delta L}{\lambda_P}$. The interferometer stabilization, at a desired phase set-point $\varphi_0$, is thus simply obtained by controlling the fringe position on the CCD array.


Because of environmental noise, the actual phase value $\varphi$ drifts with respect to the desired set point $\varphi_0$ (red continuous line in Fig \ref{Fig1} (b)) by some random quantity $\varphi - \varphi_0 = \delta$, leading to a shifted interference pattern on the CCD sensitive region (blue dotted line in Fig \ref{Fig1} (b)).  
We define an error function as follows:
\begin{equation}
\label{error}
\mathcal{E}\left(\delta\right)=\int f\left(\vartheta,\varphi\right) g\left(\vartheta,\varphi_0\right) \text{d}\vartheta = 4 A \sin\left(\delta\right)
\end{equation}
where $g\left(\vartheta,\varphi_0\right)= \text{sign}\left(\cos\left(\vartheta-\varphi_0\right)\right)$ is a step function (back line in Fig \ref{Fig1} (b)) with the same period and in phase quadrature with respect to the sinusoid curve on the CCD camera at the set-point phase value $\varphi_0$. The computed error value $\mathcal{E}\left(\delta\right)$ is then fed to a computer based PID control algorithm driving the piezo actuator.

The error function (\ref{error}) has many advantages. First, it is independent on the background term $b$ of eq. \ref{signal}. Most importantly, the error is \emph{independent on the set-point} $\varphi_0$ implying that all set-points are equivalent in this method. Furthermore, the error is linear and monotone around the set-point. 

\begin{figure}[t]
\centerline{\includegraphics[width=\columnwidth]{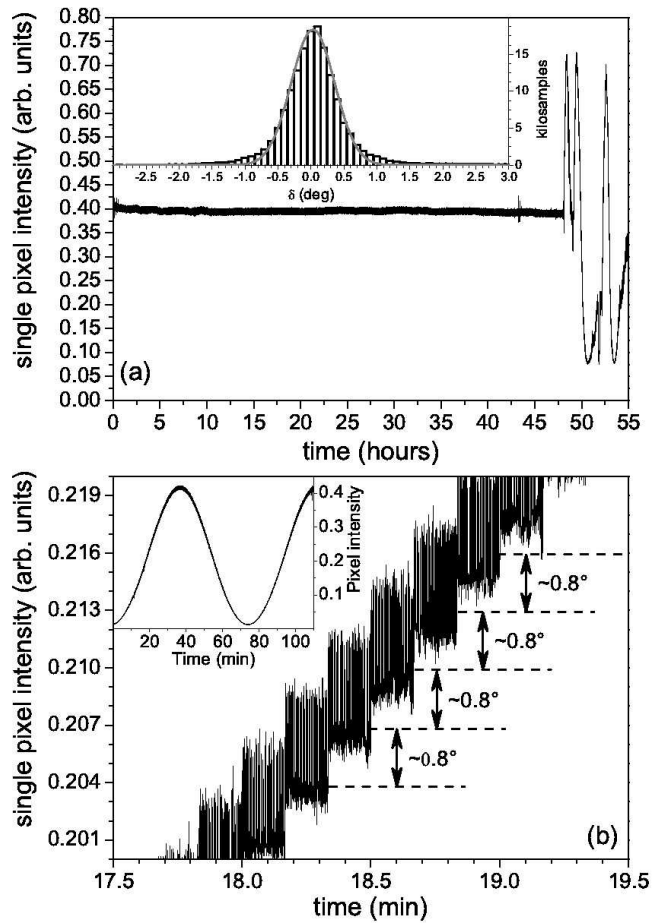}}
\caption{(a) Interferometer output with and without the feedback as a function of the acquisition time. In the inset is reported the statistical distribution of the phase errors $\delta$ during the stabilization. (b) Output of the interferometer when moved in step scan mode one pixel at time. The angle of the mirror M1 is such that one pixel on the CCD camera corresponds to about 0.8$^\circ$. }
\vspace{-15pt}
\label{Fig2}
\end{figure}

To test the method, we built a Michelson interferometer as shown in Fig. 1. A He:Ne laser ($\lambda_P=$632.8 nm, beam expanded to a diameter of 10 mm) is used as reference. The moving mirror M1 is positioned using a piezo actuator (Attocube series 101) with resolution better than 1 nm, and the CCD array is a linescan camera with 7 $\mu$m pixels. The interferometer is built on an optical table without thermal stabilization, and it is enclosed in a plastic case to provide insulation from air currents in the room. The first property we want to test is stability: in Fig 2 (a) we plot the intensity, measured on a single pixel of the camera, corresponding to $\varphi_0=0^\circ$ and $\vartheta=180^\circ$. The interferometer was kept stable for two full days, then the feedback turned off for comparison. The data collected during the stabilized run (at the rate of 40 samples per second) were statistically analyzed and the distribution of the values of $\delta$ (obtained from the data inverting eq. (\ref{signal})) is shown in the inset of Fig. 2 (a): the data can be fitted with a gaussian distribution (gray line in the inset) with average $0^\circ \pm 0.1^\circ$ and full width at half maximum of $0.7^\circ \pm\ 0.1^\circ$. The FWHM corresponds to $\sim 0.8$ nm, corresponding to the positioning precision of the piezo actuator. 

Aside from stability, another important feature of a stabilization method is positioning accuracy: to test it, we have modified the tilt angle of mirror M2 so that a full fringe took $\sim$500 pixels on the CCD. In this case a single pixel corresponds to an angle of 360$^\circ$/500=0.8$^\circ$, so that it is comparable to the measured piezo precision (Fig. \ref{Fig2} (a)). We then proceeded to stabilize the interferometer varying the set-point every 20 seconds by shifting the step function one pixel each time: the intensity measured is shown in Fig \ref{Fig2} (b); the inset shows the full dataset, a variation of more than 360$^\circ$, while the main figure shows a detail taken around half the maximum intensity. This proves the interferometer can be set at an arbitrary position with a precision limited by the piezo actuator (the step is comparable to the positioning noise, as expected).

A potential drawback of this method is that an intentional slight misalignment of the interferometer is required to produce the desired fringe pattern on the CCD camera. However, most optical applications, including metrology, holography and quantum optics, require the wavefront of the controlled beam to have the same phase all over the entire beam cross section. This problem can however be simply overcome by separating the reference beam from the controlled beam as shown in Fig. \ref{Fig1} (a). In our case we have used a cw laser at a wavelength $\lambda_M=$1550 nm, but this method can be applied to whatever wavelength compatible with the optics. In order to achieve a constant spatial profile on the IR beam cross section, we employed a lens on its path, so that the mirror M2 is in the focal plane of the lens. Due to the Fourier transform properties of the lens, the incident and reflected IR beams remain parallel, regardless of the tilt on M2. The phase uncertainty of the IR beam depends on the diffraction limit of the lens.

\begin{figure}
\centerline{\includegraphics[width=\columnwidth]{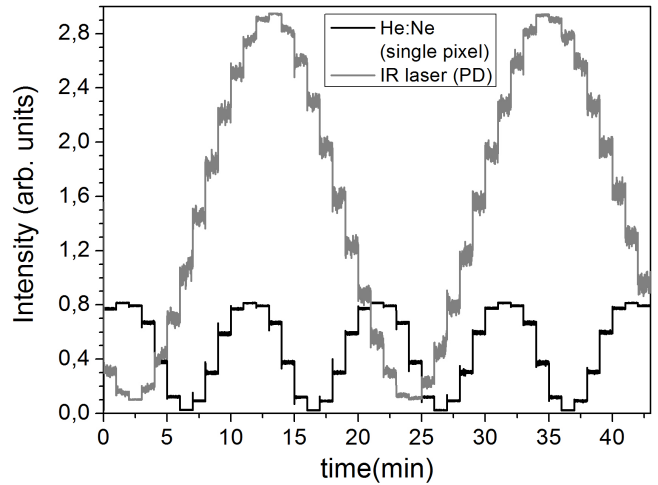}}
\caption{Step scan measure in the double beam configuration. The black line represents the intensity of the He:Ne control laser on a single pixel of the CCD camera, the gray line represents the output of the IR beam, with constant phase over its cross section area, measured on a photodetector.}
\vspace{-15pt}
\label{Fig3}
\end{figure}

The He:Ne laser can then act as a probe to set the phase of the IR beam, as shown in Fig. \ref{Fig3}. The fringe visibility on the IR beam as measured on the photodiode exceeds 94$\%$, showing that the phase is extremely homogeneous within its diameter. This measure also shows that the method can move the interferometer continuously  for angles greater than 360$^\circ$: the set-point phase $\varphi_0$ at the He:Ne frequency is shifted by more than two periods while the phase for the IR laser changes by 360$^\circ$. This is due to the linearity of the error of eq. (\ref{error}) and its independence on the set-point.

In summary, we have developed a novel method to control and stabilize an optical interferometer at an arbitrary phase. We have shown this method can be employed to position a piezoelectric actuator to its sub-nanometric limit. These results could be of importance for applications where nanopositioning accuracy is essential, as well as for extremely precise optical measurements in metrology \cite{metrology1,metrology2}, near field scanning microscopy in dispersive materials \cite{Gersen} and quantum optics experiments such as homodyne detection in quantum state tomography \cite{Smithey} or interferometric measurements for entanglement detection \cite{Marikic}.

This work was supported by CNISM funding through the INNESCO project "PcPol", by MIUR funding through the FIRB ``Futuro in Ricerca" project RBFR08XMVY and from the foundation Alma Mater Ticinensis. We acknowledge Stefano Azzini and Marco Liscidini for careful reading of the manuscript.

\pagebreak

\end{document}